\documentstyle[12pt]{article}

\begin{document}
\def\thebibliography#1{\section*{REFERENCES\markboth
 {REFERENCES}{REFERENCES}}\list
 {[\arabic{enumi}]}{\settowidth\labelwidth{[#1]}\leftmargin\labelwidth
 \advance\leftmargin\labelsep
 \usecounter{enumi}}
 \def\newblock{\hskip .11em plus .33em minus -.07em}
 \sloppy
 \sfcode`\.=1000\relax}
\let\endthebibliography=\endlist

\hoffset = -1truecm
\voffset = -2truecm


\title{\large\bf
MarKov-Yukawa Transversality On Covariant Null Plane:
Pion Form Factor, Gauge Invariance And Lorentz Completion
}
\author{
{\normalsize\bf
A.N.Mitra \thanks{e.mail: (1)ganmitra@nde.vsnl.net.in;
(2)anmitra@csec.ernet.in}
}\\
\normalsize 244 Tagore Park, Delhi-110009, India.}

\date{6 May 1999}

\maketitle

\begin{abstract}
The Markov-Yukawa Transversality Principle (MYTP) on a 2-body Bethe-Salpeter
kernel is formulated on a covariant Null Plane (NP) to reconstruct
the 4D BS wave function for 2 fermion quarks in terms of 3D entities
that satisfy a 3D BSE. This result is the null-plane counterpart of a 3D-4D
interconnection for the 2-body BS wave functions  found earlier by imposing
MYTP covariantly in the instantaneous rest frame (termed CIA) of the
composite. This formulation  yields a 3D BSE which is formally identical
to its Covariant Instantaneity form, thus fully preserving its spectral
results, while ensuring full covariance. More importantly, the reconstructed
4D vertex functions in the covariant null-plane ensure that 4D quark-loops
are now free from ill-defined time-like momentum integrations
(which had plagued the earlier CIA vertex functions), while a simple
prescription of `Lorentz completion' in the new description yields a
manifestly Lorentz-invariant result.This is illustrated for the pion and
kaon form factors with full QED gauge-invariance, showing a $k^{-2}$
behaviour at large $k^2$, and `correct' slopes at small  $k^2$. This method
is compared with the Kadychevsky-Karmanov light-front formalism. 

\end{abstract}

\newpage


\section{Introduction: Markov-Yukawa Transversality}

        For a {\it relativistic} 2-body problem, the historical issue of
3D reduction from a 4D BSE has been in the forefront of its physics from the
outset [1-3]: Instantaneous approximation [1]; Quasi-potential approach [2];
variants of on-shellness of propagators [3]. In all these methods the
starting BSE is 4D in all details, {\it {including its kernel}}, but the
associated propagators are manipulated in various ways to reduce the 4D
BSE to a 3D form as a fresh starting point, giving up its original 4D form,
more in conformity with the Weinberg 3D spirit [4]. Kadychevsky [5] and
Karmanov [6] have given it a more formal shape with on-shell propagators
and spurions [6],  making up a covariant light-front formulation which has
been recently reviewed by Carbonell et al [7]. (This paper will be termed
KK [7] in the text).
\par
        An alternative approach of more recent origin [8,9] is based on the
Markov-Yukawa Transversality Principle (MYTP) [10] wherein a
Lorentz-covariant
3D support is postulated at the outset to the pairwise BSE kernel $K$ by
demanding that it be a function of only ${\hat q}_\mu = q_\mu-q.PP_\mu/P^2$,
implying ${\hat q}.P \equiv 0$; but the propagators are left untouched in
their original 4D forms.  This is somewhat complementary to the approaches
[1-3] (propagators manipulated but kernel left untouched), so that the
resulting equations [8-9] look unfamiliar vis-a-vis these [1-3], but it has
the advantage of allowing a {\it {double track}} use of both 3D and 4D BSE
forms via their interlinkage. Indeed what distinguishes the Covariant
Instaneity Ansatz (CIA) [8] from the more familiar 3D reductions of the BSE
[1-3] is its capacity for a 2-way connection: an exact 3D BSE reduction, and
an equally exact reconstruction of the original 4D BSE form without extra
charge [8]: the former to access the observed O(3)-like spectra [11], and
the
latter to give transition amplitudes as 4D quark loop integrals [8]. [In the
corresponding approach of the Pervushin Group too [9], this twin feature was
also present, but seemingly unnoticed]. In contrast the more orthodox
methods
[1-3] give only a one-way connection,  viz., $4D\rightarrow 3D$ reduction,
but {\it not} vice versa. This unique feature of MYTP [10], providing a
2-way
3D-4D interconnection in the BSE structure, has somehow remained unnoticed
in the literature, despite a demonstration [8] of its existence.

\subsection{Physics of MYTP on Null-plane}
\par
        The MYTP [10] controlled BSE, termed 3D-4D-BSE in the following, of
course needs supplementing by physical ingredients to define the BSE kernel,
much as a Hamiltonian needs a properly defined `potential'. However its
canvas
is broad enough to accommodate a wide variety of kernels which must in turn
be governed by independent physical principles. In this respect, the
orthodox
view (which we adopt) is to keep close to the traditional 4D BSE-cum-SDE
methods [12] of Dynamical Breaking of Chiral Symmetry ($DB{\chi}S$) a la NJL
[13] whose basic feature of chiral symmetry breaking survives a (space-time
extended) 4-quark interaction mediated by vector exchange [14] as a
generalized
$DB{\chi}S$ mechanism to generate a mass-function $m(p)$ via Schwinger-Dyson
equation (SDE), which accounts for the bulk of the constituent mass of
$ud$ quarks via Politzer additivity [15]. Indeed, the BSE-SDE formalism
[12-15] can be simply adapted [16] to the hybrid 3D-4D-BSE form [8] which
produces 3D spectra of both hadron types [17] under a common parametrization
for the gluon propagator, with a self-consistent SDE determination [16] of
the constituent mass.
\par
        One disadvantage of MYTP [10] for 3D-4D interconnection [8] achieved
under  covariant instantaneity (CIA) in the composite's rest frame [8] is
the
ill-defined nature of 4D loop integrals which acquire time-like momentum
components in the exponential/gaussian factors associated with the different
vertex functions due to a `Lorentz-mismatch' among the rest-frames of the
participating hadronic composites. This is especially so for triangle loops
and above, such as the pion form factor, while 2-quark loops [18] just
escape this pathology. This problem was not explicitly encountered in the
null-plane ansatz (NPA) [19] in an earlier study of 4D triangle loop
integrals, but the NPA approach was criticized [20] on grounds of
non-covariance. The CIA approach [8] which makes use of MYTP [10,9], was
an attempt to rectify the Lorentz covariance defect, but the presence of
time-like components in the gaussian factors inside triangle loop integrals,
e.g., in the pion form factor [21], impeded further progress.
\par
        In this paper we wish to explore if it is possible to ensure formal
covariance without having to encounter the time-like components in the
(gaussian) wave functions inside the 4D loop integrals.  This paper is an
attempt to show that by extending the Transversality Principle [10] from the
covariant rest frame of the (hadron) composite [8-10], to a {\it
{covariantly
defined}} null-plane (NP) both these features can be preserved, so that even
the old-fashioned null-plane framework [19,22] for 3D-4D interconnection can
be given a covariant "look". In this respect, we shall also compare the
present `TP-NP' method with other covariant NP approaches [23], especially
KK [7], which has some obvious similarities with the earlier (old-fashioned)
NPA formulation [22].

\subsection{Scope of the Paper}
\par
        This paper has a 4-fold objective: i) To formulate MYTP [10] on a
covariant null-plane (NP), i.e., demand that the BSE kernel $K$ for pairwise
interaction is a function of relative momentum $q$ {\it transverse} to the
NP,
just as in CIA [8] the Transversality is w.r.t. the composite 4-momentum
$P_\mu$; ii) to show that the reduced 3D BSE has formally the same structure
as in the corresponding CIA [8] case, so that with the same parametrization
for $K$, the spectral predictions [17] remain unchanged; iii) to show that
the reconstructed 4D wave function no longer suffers a `Lorentz-mismatch'
with other such functions involved in a loop integral (which, in CIA [8,21],
is the main cause for the appearence of time-like components in the gaussian
form factors); iv) to illustrate the detailed procedure by working out the
pion-form factor under QED gauge invariance and a simple prescription of
`Lorentz-completion' to obtain an explicitly Lorentz- and gauge-invariant
quantity which shows the desired $k^{-2}$ behaviour at high $k^2$. For a
better perspective, we shall also offer a critical comparison of this
procedure with  the covariant  NP approach of KK [7].
\par
        After some preliminaries on the definition of a covariant
null-plane,
Sec.2 employs MYTP [10] for the BS kernel $K$ for spinless quarks to
define its structure on such a null-plane, by close analogy to the CIA
formulation [8], and outlines the derivation of the 3D BSE, as well as an
explicit reconstruction of the 4D wave function in terms of 3D ingredients,
in which the 3-momentum is  ${\hat q} = (q_\perp, q_3)$, where the
third component emerges as a $P$-dependent one lying in the null plane (NP).
Sec.3 is a self-contained derivation of the 3D BSE for a fermion pair, in
which the full structure of the non-perturbative gluonic kernel [16] is
redefined in the NP language, so as to bring out a strong similarity to the
corresponding CIA equation [24], to justify identical predictions on the
spectroscopy front [17]. With this covariant NP-oriented 3D-4D formulation,
Sec.4 outlines the calculation of the P-meson form factor for unequal
mass kinematics in a fully gauge invariant manner, including correction
terms
arising from QED gauge invariance, and illustrating the techniques of
`Lorentz-completion' to obtain an explicitly Lorentz invariant quantity.
As a check on the consistency of the formalism, the expected $k^{-2}$
behaviour of the pion form factor at high $k^2$ is realized. Sec.5 concludes
with a short discussion, a summary of this method in retrospect, plus a
comparison with KK [7] as a prototype for other NP approaches [23].
Some calculational details on the triangle-loop integral for the P-meson
form factor are given in Appendix A.


\section{3D-4D BSE Formalism on Covariant Null-Plane}

As a preliminary to defining a 3D support to the BS kernel on the
null-plane (NP), on the lines of CIA [8], a covariant NP orientation may be
represented by the 4-vector $n_\mu$, as well as its dual ${\tilde n}_\mu$,
obeying the normalizations $n^2 = {\tilde n}^2 =0$ and $n.{\tilde n} = 1$.
In the standard NP scheme (in euclidean notation), these quantities
are $n=(001;-i)/\sqrt{2}$ and ${\tilde n}=(001;i)/\sqrt{2}$, while the two
other perpendicular directions are collectively denoted by the subscript
$\perp$ on the concerned momenta. We shall try to maintain the
$n$-dependence
of various momenta to ensure explicit covariance; and to keep track of
the old-fashioned NP notation $p_{\pm} = p_0 \pm p_3$, our covariant
notation
is normalized to the latter as  $p_+ = n.p \sqrt{2}$;
$p_- = -{\tilde n}.p \sqrt{2}$, while the perpendicular components continue
to be denoted by $p_{\perp}$ in both notations.
\setcounter{equation}{0}
\renewcommand{\theequation}{2.\arabic{equation}}
\par
        For the various quantities (masses, momenta, etc) we shall stick to
the notation of [8] without explanation, except when new features arise.
For the relative momentum $q={\hat m}_2 p_1-{\hat m}_1 p_2$, the fourth
component to be eliminated for obtaining a 3D equation, is proportional
to $q_n \equiv {\tilde n}.q$, as the NP analogue [22] of $P.qP/P^2$ in
CIA [8], where $P=p_1+p_2$ is the total 4-momentum of the hadron. However
the quantity $q - q_n n$ is still only $q_\perp$, since its square is
$q^2 - 2 n.q{\tilde n}.q$, as befits $q_\perp^2$ (readily checked against
the
`special' NP frame). We still need a third component $p_3$, for which a
first guess is $zP$, where $z=n.q/n.P$. And for calculational convenience
we shall need to (temporarily) invoke the `collinear frame' which amounts
to $P_\perp.q_\perp = 0$, a restriction which will be removed later by a
simple prescription of `Lorentz completion'. Unfortunately the definition
${\hat q}_\mu$ = $(q_{\perp\mu}, zP_\mu)$ does not quite fit the bill for a
covariant 3-vector, since a short calculation shows again that ${\hat q}^2$
=$q_\perp^2$. The correct definition turns out to be $q_{3\mu} = z P_n
n_\mu$,
where $P_n$=$P.{\tilde n}$, giving ${\hat q}^2$ = $q_\perp^2 + z^2M^2$, as
required. We now collect the following definitions/results:
\begin{eqnarray}
 q_\perp &=& q-q_n n; \quad {\hat q}=q_\perp+ z P_n n; \quad z=q.n/P.n;
 \quad P^2 = -M^2; \\ \nonumber
 q_n,P_n &=& {\tilde n}.(q,P);{\hat q}.n = q.n; \quad {\hat q}.{\tilde n} =
0;
\quad P_\perp.q_\perp = 0;   \\ \nonumber
     P.q &=& P_n q.n + P.n q_n; \quad P.{\hat q} = P_n q.n; \quad
{\hat q}^2 = q_\perp^2 + M^2 x^2
\end{eqnarray}

\subsection{3D-4D BSE on Cov. NP : Spinless Quarks}

        We now proceed to derive the reduced 3D BSE (wave-fn $\phi$) from
the 4D BSE with spinless quarks (wave-fn $\Phi$) when its kernel $K$ is
{\it decreed} to be independent of the component $q_n$, i.e.,
$K=K({\hat q},{\hat q}')$, with ${\hat q}$ = $(q_\perp,  P_n n)$,
in accordance with the TP [10] condition imposed on the null-plane (NP).
The 4D BSE with such a kernel is, c.f., [8,9]:
\begin{equation}
i(2\pi)^4 \Phi(q) = {\Delta_1}^{-1} {\Delta_2}^{-1}  \int d^4q'
K({\hat q},{\hat q'}) \Phi(q')
\end{equation}
where  $\Delta_i$ = ${p_i}^2 + {m_i}^2$; $m_i$ = quark mass; $(P^2 = -M^2)$;
and $d^4 q$ = $d^2 q_\perp dq_3 dq_n$. Now define a 3D wave-fn
$\phi({\hat q})$ = $ \int  d{q_n}\Phi(q)$, and use this result on the
RHS of (2.2) to give
\begin{equation}
i(2\pi)^4 \Phi(q) = {\Delta_1}^{-1} {\Delta_2}^{-1}
\int d^3 q' K({\hat q},{\hat q}')\phi({\hat q}')
\end{equation}
Now integrate both sides of eq.(2.3) w.r.t. $dq_n$ to give a 3D BSE
in the variable ${\hat q}$:
\begin{equation}
(2\pi)^3 D_n({\hat q}) \phi({\hat q}) =  \int d^2{q_\perp}'d{q_3}'
K({\hat q},{\hat q}') \phi({\hat q}')
\end{equation}
where the function $D_n(\hat q)$, is defined, analogously to CIA [8], by
\begin{equation}
\int d{q_n}{\Delta_1}^{-1} {\Delta_2}^{-1} = 2{\pi}i D_n^{-1}({\hat q})
\end{equation}
and may be obtained by standard NP techniques [22] (Chaps 5-7) as follows.
In the $q_n$ plane, the poles of $\Delta_{1,2}$ lie on opposite sides of
the real axis, so that only {\it one} pole will contribute at a time. Taking
the $\Delta_2$-pole, which gives
\begin{equation}
2q_n = -{\sqrt 2} q_- = {{m_2^2 + (q_\perp-{\hat m}_2P)^2}
 \over {{\hat m}_2 P.n - q.n}}
\end{equation}
the residue of $\Delta_1$ works out, after a routine simplification, to
just $2P.q = 2P.n q_n+2P_n q.n$, after using the collinearity condition
$P_\perp.q_\perp = 0$ from (2.1). And when the value (2.6) of $q_n$ is
substituted in (2.5), one obtains (with $P_n P.n = -M^2/2$):
\begin{equation}
D_n({\hat q}) = 2P.n ({\hat q}^2 -{{\lambda(M^2, m_1^2, m_2^2)} \over
{4M^2}});
\quad {\hat q}^2 = q_\perp^2 + M^2 x^2; \quad x = q.n/P.n
\end{equation}
Now a comparison of (2.2) with (2.4) relates the 4D and 3D wave-fns:
\begin{equation}
2{\pi}i \Phi(q) = D_n({\hat q}){\Delta_1}^{-1}{\Delta_2}^{-1} \phi({\hat q})
\end{equation}
which is valid near the bound state pole. The BS vertex function now becomes
$\Gamma = D_n \times \phi/(2{\pi}i)$, analogously to the CIA result [8].
This
result, though dependent on the NP orientation, is nevertheless formally
{\it covariant}, and meets the `covariance' criticism [20] of the earlier
NPA formulation [19,22], where an identical result for $D_+$ had been found.
For fermion quarks, the $q{\bar q}$ wave function $\Psi$ has formally the
same structure as (2.8), except that $D_F(p_i)=-i{\Delta(p_i)}^{-1}$ is
replaced by $S_F(p_i)$, and the vertex function $\Gamma$ has an extra factor
$\gamma_5$ for pseudoscalar, $i\gamma_\mu$ for vector, etc meson [8,24].
\par
        A few comments are now in order on a comparison of the KK [7] vs
the present covariant formulation of NP dynamics [19,22]. Both are
`covariantly' dependent on the orientation $n_\mu$ of the NP, i.e., have
certain $n$-dependent 3-scalars, in addition to genuine 4-scalars. Note
also the formal identity of (2.4) (and (2.7)) with KK's eq.(3.48) [7]. Our
$n_\mu$ is KK's $\omega_\mu$, but we have an {\it independent} 4-vector
${\tilde n}_\mu$ which has a dual interplay with $n_\mu$ in the above
formulation, but without a counterpart in KK [7]. Secondly our manifestly
covariant 4D formulation needs no 3-vector like ${\bf n}$ [7], nor explicit
Lorentz transformations. As to the `angular condition' [25] discussed in KK,
we have not (in our Cov. NP formalism) had to make any special effort to
satisfy this, since the very appearance of the `effective' 3-vector
${\hat q}_\mu$ in the 3D BSE in a rotationally invariant manner is an
automatic guarantee (in the sense of the `proof of the pudding') of the
satisfaction of this condition [25] without further assumptions.
\par
        Our 3D-4D (hybrid) BSE formulation (which allows for off-shell
momenta) has no further need for spurions which KK [7] require to make up
for energy-momentum balance since all their {\it physical} momenta are
on-shell. Nor are fresh Feynman techniques with spurions [7] needed, since
normal 4D techniques apply directly to the 3D-4D formalism per se [19,22].
Finally, to rid the {\it physical} amplitudes of $n_\mu$-dependent terms
in the external (hadron) momenta, after integration over the internal loop
momenta, we shall use a simple technique of `Lorentz-completion' (to be
illustrated in Sec.4 for the pion form factor calculation) as an alternative
to other NP prescriptions [7,23] to remove ${\bf n}$-dependent terms.
\par
        A more succinct comparison with other null-plane approaches concerns
the inverse process of {\it reconstruction} of the 4D hadron-quark vertex,
eq.(2.8), which does not seem to have a counterpart in these, e.g., KK [7],
or
the Wilson group [23], which are basically 3D oriented. Thus the nearest
analogue of this in KK [7] is to express the 3D NP wave function in terms of
the 4D BS wave function (see eq.(3.58) of KK [7]), but {\it {not vice
versa}}.
This illustrates the problem of "loss of Hilbert space information" inherent
in such a process of reconstruction, as discussed recently in the context of
the $qqq$ problem [26]. The TP [10] is a big help in this regard: For the
two-body $q{\bar q}$ case, the inversion is {\it exact} [8] (this is a sort
of degenerate situation). However for 3 or more `bodies', even TP [10] has
its limitations, since some additional assumptions are needed to fill the
information gap; (see [26] for more details).

\section{3D Fermion BSE on Covariant Null Plane}

In this Section we shall collect under one head the covariant counterparts
of the main features of an earlier CIA [8,16] and (old-fashioned) NP [22]
3D formulations, and indicate their ramifications on Spectroscopy. We
stress at the outset that the Transversality Principle (MYTP), when applied
to the Covariant null-plane (CNP), gives formally the same structure of the
3D BSE as the application of MYTP gives under CIA [8,16], or even under
the old-fashioned non-covariant NPA [22].
\setcounter{equation}{0}
\renewcommand{\theequation}{3.\arabic{equation}}
\par
        The 4D BSE for fermionic quarks under a gluonic (vector-type)
interaction kernel with 3D support has the standard form [24]:
\begin{equation}
i(2\pi)^4 \Psi(P,q) = S_{F1}(p_1) S_{F2}(p_2) \int d^4 q'
K({\hat q}, {\hat q}') \Psi(P,q'); \quad K= F_{12} i\gamma_\mu^{(1)}
i\gamma_\mu^{(2)} V({\hat q},{\hat q}')
\end{equation}
where $F_{12}$ is the color factor $\lambda_1.\lambda_2/4$ and the $V$-
function expresses the scalar structure of the gluon propagator in the
perturbative (o.g.e.) plus non-perturbative regimes, whose full structure
(as employed in actual calculations) [22,17] is collected as under, using
the simplified notations $k$ for $q-q'$, and $V({\hat k})$ for the $V$ fn:
\begin{equation}
V({\hat k})=4\pi\alpha_s/{\hat k}^2+{3 \over 4}\omega_{q{\bar q}}^2
 \int d{\bf r} [r^2 {(1+4A_0 {\hat m}_1{\hat m}_2 {M_>}^2 r^2)}^{-1/2}
 -C_0/\omega_0^2] e^{i{\hat k}.{\bf r}};
\end{equation}
\begin{equation}
\omega_{q{\bar q}}^2 = 4M_> {\hat m}_1 {\hat m}_2 \omega_0^2
\alpha_s({M_>}^2);
\quad \alpha_s(Q^2)={{6\pi} \over {33-2n_f}} {\ln (M_>/\Lambda)}^{-1};
\end{equation}
\begin{equation}
{\hat m}_{1,2} = [1 \pm (m_1^2 - m_2^2)/M^2]/2; \quad
M_> = Max (M, m_1+m_2); \quad C_0 = 0.27;\quad A_0=0.0283
\end{equation}
And the values of the basic constants (all in $MeV$) are [17]
\begin{equation}
\omega_0= 158 ;\quad m_{ud}=265 ;\quad m_s=415 ; \quad m_c=1530 ;
\quad m_b = 4900.
\end{equation}
\par
        The BSE form (3.1) is however not the most convenient one for wider
applications in practice, since the Dirac matrices entail several coupled
integral equations. Indeed, as noted long ago [24], a considerable
simplification is effected by expressing them in `Gordon-reduced' form,
(permissible on the quark mass shells, or better on the surface $P.q = 0$),
a step which may be regarded as a fresh starting point of our dynamics, in
the
sense of an `analytic continuation' of the $\gamma$- matrices to `off-shell'
regions (i.e., away from the surface $P.q =0$). Admittedly this constitutes
a conscious departure from the original BSE structure (3.1), but such
technical modifications are not unknown in the BS literature [27] in the
interest of greater manoeuvreability, without giving up the essentials, in
view of the  "effective" (and not fundamental) nature of the BS kernel.
\par
        The `Gordon-reduced' BSE form of (3.1) is given by [24]
\begin{equation}
\Delta_1 \Delta_2 \Phi(P,q) = -i(2\pi)^{-4} F_{12} \int d^4q' V_\mu^{(1)}
V_\mu^{(2)} V({\hat q},{\hat q}') \Phi(P,q');
\end{equation}
where the connection between the $\Psi$- and $\Phi$- functions is
\begin{equation}
\Psi(P,q) = (m_1-i\gamma^{(1)}.p_1)(m_2+i\gamma^{(2)}.p_2) \Phi(P,q); \quad
 p_{1,2} = {\hat m}_{1,2} P \pm q
\end{equation}
\begin{equation}
V_\mu^{(1,2)}= \pm 2m_{1,2} \gamma_\mu^{(1,2)}; \quad
V_\mu^{(i)}=p_{i\mu}+p_{i\mu}'+i \sigma_{\mu\nu}^{(i)}(p_{i\nu}-p_{i\nu}')
\end{equation}
\par
        Now to implement the Transversality Condition [8-10] for the entire
kernel of eq.(3.6), all time-like components $q_n,q_n'$ in the product
$V^{(1)}.V^{(2)}$, as defined in eq.(2.1), must first be {\it dropped}.
Substituting from (3.8) and simplifying  gives
\begin{equation}
(p_1+p_1').(p_2+p_2') = 4{\hat m}_1{\hat m}_2 P^2 -({\hat q}+{\hat q}')^2
-2({\hat m}_1-{\hat m}_2) P.(q+q') + ``spin-Terms'';
\end{equation}
\begin{equation}
``Spin Terms'' = -i(2{\hat m}_1P+{\hat q}+{\hat
q}')_\mu\sigma_{\mu\nu}^{(2)}
{\hat k}_\nu+i(2{\hat m}_2P-{\hat q}-{\hat q}')_\mu \sigma_{\mu\nu}^{(1)}
{\hat k}_\nu + \sigma_{\mu\nu}^{(1)} \sigma_{\mu\nu}^{(2)}
\end{equation}
This is identical to eq.(7.1.9) of Ref.[22], via the formal correspondence
${\hat q}_\mu$ $\Leftrightarrow$ $q_{\perp \mu}+q.n P_n n_\mu/P.n$, where
the covariance is now explicit. The 3D reduction of eq.(3.4) now goes
through
exactly as in sec.2.1, so that without further ado, the full structure of
the
3D BSE  can be literally taken over from Ref.[22]-Chap 7 (derived under
non-covariant NPA). In particular, for harmonic confinement,  obtained by
dropping the $A_0$ term in the `potential' $U(r)$ of (3.4) (a very good
approximation for {\it light} ($ud$) quarks), the 3D BSE works out as
\begin{equation}
D_n \phi({\hat q})={P_n \over M}\omega_{q{\bar q}}^2 {\tilde D}({\hat q})
\phi({\hat q});
\end{equation}
\begin{equation}
{\tilde D}({\hat q})= 4{\hat m}_1{\hat m}_2 M^2 ({\nabla}^2+C_0/\omega_0^2)
+ 4{\hat q}^2{\nabla}^2+8{\hat q}.{\nabla}+18
- 8{\bf J}.{\bf S}+(4C_0/\omega_0^2){\hat q}^2
\end{equation}
Note that the covariance is again manifest, since both sides are
proportional
to $P_n$, as checked from  eq.(2.7) for $D_n$. Since one of the objects of
this exercise is to strengthen the mathematical foundations of results
[22,17]
that have already made successful contact with data, we may simply refer
to [22,17] for the details of spectroscopic, etc predictions. For the sake
of
algebraic completeness however, we record the (gaussian) parameter $\beta$
of
the 3D wave function $\phi({\hat q})$ = $exp(-{\hat q}^2/{2 \beta^2})$,
which
is the solution of (3.11) for a ground state hadron [22,16-18]:
\begin{equation}
\beta^4 = {{8{\hat m}_1^2{\hat m}_2^2 M_>^2 \omega_0^2 \alpha(M_>^2)}
\over {[1 - 8 C_0 {\hat m}_1 {\hat m}_2 \alpha_s(M_>^2)] <\sigma>}};\quad
<\sigma>^2 = 1 + 24 A_0 ({\hat m}_1{\hat m}_2 M_>)^2 /\beta^2
\end{equation}
Note that $\beta$ is {\it independent} of the null-plane orientation
$n_\mu$.
(For an $L$-excited hadron wave function, see [19,22]). The full 4D BS wave
function $\Psi(P,q)$ in a 4x4 matrix form [22] is then reconstructed from
(3.6-7) as [19,22,8]
\begin{equation}
\Psi(P,q)= S_F(p_1) \Gamma({\hat q}) \gamma_D S_F(-p_2); \quad
\Gamma({\hat q}) = N_n(P) D_n ({\hat q})\phi({\hat q})/{2i\pi}
\end{equation}
where $\gamma_D$ is a Dirac matrix which equals $\gamma_5$ for a P-meson,
$i\gamma_\mu$ for a V-meson, $i\gamma_\mu \gamma_5$ for an A-meson, etc.
$N_n(P)$ represents the hadron normalization which is derived in Sec.4.
\par
        We end this section with the remark that the TP [10] approach on the
covariant null plane has yielded a wave function whose form resembles one of
the varieties obtained in the old-fashioned NPA treatment [19,22] that had
been termed the ``on-shell'' variety. However, two other varieties
encountered
in the old-fashioned treatment, viz., ``off-shell'' and ``half-off-shell'',
have not found a natural counterpart in this manifestly covariant treatment.

\section{P-Meson E.M. Form Factor for Unequal masses}

The pion form factor has through the ages been a good laboratory for
subjecting theoretical models and ideas on strong interactions to
observational test. Among the crucial parameters are the squared radius
$<r_{expt}^2>$ = $0.43 \pm .014 fm^2$ [28a], and the scaled form factor
at high $k^2$, viz., $k^2 F(k^2) \approx 0.5 \pm 0.1 GeV^2$ [28b] which
represent important check points for theoretical candidates such as
QCD-sum rules [29], Finite Energy sum rules [30], perturbative QCD [31],
covariant null-plane approaches [7, 23], 4D SDE-BSE methods [12], including
Euclidean SDE [32], etc. An important issue in this regard concerns the
interface of perturbative and non-perturbative QCD regimes. Indeed
it has been suggested on the basis of certain model studies in terms of
light-front wave functions with specific combinations of longitudinal and
transverse components [33], that non-perturbative effects probably persist
up to the highest $k^2$, as deduced from the relative importance of the
transverse component vis-a-vis the longitudinal, for all  $k^2$.
\par
        While such a conclusion [33] sounds apriori reasonable, the
demonstration of any intimate connection between the transverse and
longitudinal components of the (light front) wave function is the task
of a more detailed dynamical theory (such as an effective Lagrangian)
than can be captured by such ad hoc but intuitive ansatze [33]. From this
angle a dynamical model such as the covariant NP formalism developed in
Secs 2-3 (pre-calibrated to spectroscopic details [17]) would appear to
be a more promising canditate, like [7,23,29-32], for addressing such
dynamical issues. Indeed the pion e.m. form factor was already worked
out in the old NPA formalism [19], with very reasonable results, but we
outline below a fresh, self-contained `covariant' derivation valid for
unequal masses of the quark constutuents, one in which  issues of gauge
invariance as well as of `Lorentz-completion' will receive particular
attention, with a view to check (among other things) on the expected
$k^{-2}$  behaviour at high $k^2$, and the `slope' at low $k^2$.

\subsection{4D Triangle Loop Integral for $F(k^2)$}

\setcounter{equation}{0}
\renewcommand{\theequation}{4.\arabic{equation}}

        Using the two diagrams (figs.1a and 1b) of ref.[19], and in the
same notation, the Feynman amplitude for the $h \rightarrow h'+ \gamma$
transition contributed by fig.1a ( quark $2$ as spectator is given by [19]
\begin{equation}
2{\bar P}_\mu F(k^2)= 4(2\pi)^4 N_n(P)N_n(P') e{\hat m}_1 \int d^4
T_\mu^{(1)}
{{D_n({\hat q})\phi({\hat q})D_n({\hat q}')\phi({\hat q}')} \over
{\Delta_1 \Delta_1' \Delta_2}} + [1 \Rightarrow 2];
\end{equation}
\begin{equation}
4T_\mu^{(1)} = Tr [\gamma_5 (m_1-i\gamma.p_1) i\gamma_\mu (m_1-i\gamma.p_1')
\gamma_5 (m_2 + i\gamma.p_2)]; \quad \Delta_i = m_i^2 + p_i^2;
\end{equation}
\begin{equation}
p_{1,2} = {\hat m}_{1,2}P \pm q; \quad p_{1,2}'={\hat m}_{1,2}P' \pm q'
\quad p_2 = p_2'; \quad P-P'=p_1-p_1'=k; \quad 2{\bar P} = P+P'.
\end{equation}
After evaluating the traces and simplifying via (4.2-3), $T_\mu$ becomes
\begin{equation}
T_\mu^{(1)}= (p_{2\mu}-{\bar P}_\mu)[{\delta
m}^2-M^2-\Delta_2] -k^2p_{2\mu}/2
+(\Delta_1-\Delta_1')k_\mu/4
\end{equation}
The last term in (4.4) is non-gauge invariant, but it does not survive the
integration in (4.1), since the coefficient of $k_\mu$, viz.,
$\Delta_1-\Delta_1'$ is antisymmetric in $p_1$ and $p_1'$, while the rest
of the integrand in (4.1) is symmetric in these two variables. Next, to
bring
out the proportionality of the integral (4.1) to ${\bar P}_\mu$, it is
necessary to resolve $p_2$ into the mutually perpendicular components
$p_{2\perp}$, $(p_2.k/k^2)k$ and $(p_2.{\bar P}/{\bar P}^2){\bar P}$, of
which the first two will again not survive the integration, the first due
to the angular integration, and the second due to the antisymmetry of
$k=p_1-p_1'$ in $p_1$ and $p_1'$, just as in the last term of (4.4). The
third
term is explicitly proportional to ${\bar P}_\mu$, and is of course
gauge invariant since ${\bar P}.k =0$. (This fact had been anticipated while
writing the LHS of (4.4)). Now with the help of the results
\begin{equation}
p_2.{\bar P}= -{\hat m}_2 M^2 -\Delta_1/4 - \Delta_1'/4; \quad
2{\hat m}_2 = 1-(m_1^2-m_2^2)/M^2; \quad {\bar P}^2=-M^2-k^2/4,
\end{equation}
it is a simple matter to integrate (4.1), on the lines of Sec.2, noting
that terms proportional to $\Delta_1 \Delta_2$ and $\Delta_1' \Delta_2$
will give zero, while the non-vanishing terms will get contributions
only from the residues of the $\Delta_2$-pole, eq.(2.5). Before collecting
the various pieces, note that the 3D gaussian wave functions $\phi,\phi'$,
as well as the 3D denominator functions $D_n, D_n'$, do {\it not} depend
on the time-like components $p_{2n}$, so that no further pole contributions
accrue from these sources. (It is this problem of time-like components of
the internal 4-momenta inside the gaussian $\phi$-functions under the CIA
approach [8], that had plagued an earlier study of the pion form factor
[21], and had to be abandoned). To proceed further, it is now convenient
to define the quantity ${\bar q}.n = p_2.n -{\hat m}_2 {\bar P}.n$ to
simplify the $\phi$- and $D_n$- functions. To that end define the symbols:
\begin{equation}
(q,q') = {\bar q} \pm {\hat m}_2 k/2; \quad  z_2 = {\bar q}.n/{{\bar P}.n};
\quad {\hat k}= k.n/{{\bar P}.n}; \quad (\theta_k, \eta_k)=1 \pm {\hat
k}^2/4
\end{equation}
and note the following results of pole integration w.r.t. $p_{2n}$ [22]:
\begin{equation}
\int dp_{2n}{1 \over {\Delta_2}} [1/{\Delta_1}; 1/{\Delta_1'};
1/(\Delta_1 \Delta_1')] = 2i\pi [1/D_n; 1/D_n'; 2p_2.n/(D_nD_n')]
\end{equation}
The details of further calculation of the form factor are given in
Appendix A. An essential result is the normalizer $N_n(P)$ of the hadron,
obtained by setting $k_\mu=0$, and demanding that $F(0)=1$. The reduced
normalizer $N_H= N_n(P)P.n/M$, which is Lorentz-invariant, is given via
eq.(A.9) by:
\begin{equation}
N_H^{-2}=2M (2\pi)^3 \int d^3{\hat q}
e^{-{\hat q}^2/\beta^2} [(1+{\delta m}^2/M^2)({\hat q}^2-\lambda/{4M^2})
+2{\hat m}_1{\hat m}_2 (M^2-{\delta m}^2)]
\end{equation}
where the internal momentum ${\hat q}=(q_\perp,Mz_2)$ is formally a
3-vector,
in conformity with the `angular condition' [25]. The corresponding
expression for the form factor is (see Appendix A):
\begin{equation}
F(k^2)= 2M N_H^2 (2\pi)^3 exp[-{(M{\hat m}_2{\hat k}/\beta)^2/{4\theta_k}}]
(\pi \beta^2)^{3/2} {\eta_k \over {\sqrt \theta_k}}{\hat m}_1 G({\hat k}) +
[1 \Rightarrow 2]
\end{equation}
where $G({\hat k})$ is defined by eqs.(A.12-13) of Appendix A.

\subsection{`Lorentz Completion' for $F(k^2)$}

The expression (4.9) for $F(k^2)$ still depends on the null-plane
orientation $n_\mu$ via the dimensionless quantity ${\hat k}$ =
$k.n/P.n$ which while having simple Lorentz transformation properties,
is nevertheless {\it not} Lorentz invariant by itself. To make it explicitly
Lorentz invariant, we shall employ a simple method of `Lorentz completion'
which is merely an extension of the `collinearity trick' empolyed at the
quark level, viz., $P_\perp.q_\perp$ = 0; see eq.(2.1). Note that this
collinearity ansatz has already become reduntant at the level of the
Normalizer $N_H$, eq.(4.8), which owes its Lorentz invariance to the
integrating out of the null-plane dependent quantity $z_2$ in (4.8). This
is of course because $N_H$ depends only on one 4-momentum (that of a
{\it {single hadron}}), so that the collinearity ansatz is exactly valid.
However the form factor $F(k^2)$ depends on {\it {two independent}}
4-momenta $P,P'$, for which the collinearity assumption is non-trivial,
since the existence of the perpendicular components cannot be wished away!
Actually the quark-level assumption $P_\perp.q_\perp$ = 0 has, so to say,
got
transferred, via the ${\hat q}$-integration in eq.(4.9), to the {\it {hadron
level}}, as evidenced from the ${\hat k}$-dependence of $F(k^2)$; therefore
an obvious logical inference is to suppose this ${\hat k}$-dependence
to be the result of the  collinearity ansatz $P_\perp.P_\perp'$ = 0 at the
hadron level. Now, under the collinearity condition, one has
\begin{equation}
P.P' = P_\perp.P_\perp'+ P.n P'.{\tilde n} + P'.n P.{\tilde n}
 = P.n P_n'+ P'.n P_n ; \quad P.{\tilde n} \equiv P_n.
\end{equation}
Therefore `Lorentz completion'(the opposite of the collinearity ansatz)
merely amounts to reversing the direction of the above equation by supplying
the (zero term) $P_\perp.P_\perp'$ to a 3-scalar product to render it a
4-scalar! Indeed the process is quite unique for 3-point functions such as
the form factor under study, although for more involved cases (e.g.,
4-point functions), further assumptions may be needed.
\par
        In the present case, the prescription of Lorentz completion is
relatively simple, being already contained in eq.(4.10). Thus since
$P,P'$= ${\bar P} \pm k/2$, a simple application of (4.10) gives
\begin{equation}
k.n k_n = +k^2; \quad {\bar P}.n {\bar P}_n = -M^2 -k^2/4;
\quad {\hat k}^2= {{4k^2} \over {4M^2+k^2}}= 4\theta_k-4=4-4\eta_k
\end{equation}
This simple prescription for ${\hat k}$ automatically ensures the 4D
(Lorentz) invariance of $F(k^2)$ at the hadron level. (It may be instructive
to compare this to the KK [7] prescription of `recognizing' the $n$-
dependent terms (unphysical) of $F(k^2)$ and then dropping them). For
more involved amplitudes (e.g., 4-point functions) too, this prescription
works fairly unambiguously, if their diagrams can be analyzed in terms of
more elementary 3-point vertices (which is often possible). We hasten to add
however that strictly speaking, a `Lorentz completion' goes beyond the
original premises of restricting the (pairwise $qq$) interaction to the
covariant null-plane (in accordance with the Transversality Principle
[8-10]),
but such `analytic continuations' are not unwarranted. Indeed a careful
scrutiny of the KK [7] prescription for implementing the angular condition
[25] would reveal the introduction of `derivative' terms, implying a tacit
enlargement of the Hilbert space beyond the null-plane (see Chap 2 of [7]).

\subsection{QED Gauge Correction to $F(k^2)$}
\par
        While the `kinematic' gauge invariance of $F(k^2)$ has already been
ensured in the main Sec.4 above, there are additional contributions to the
triangle loops - figs.1a and 1b of [19] - obtained by  inserting the photon
lines at each of the two vertex blobs instead of on the quark lines
themselves.
These terms arise from the demands of QED gauge invariance, as pointed out
by Kisslinger and Li (KL) [34] in the context of two-point functions, and
are
simulated by inserting exponential phase integrals with the e.m. currents.
However, this method (which works ideally for {\it point} interactions) is
not amenable to {\it extended} (momentum-dependent) vertex functions, and
an alternative strategy is needed, as described in the context of the e.m.
self-energy problem of the baryon [35]. We briefly recapitulate the steps.
\par
        The way to an effective QED gauge invariance lies in the
simple-minded
substitution $p_i-e_iA(x_i)$ for each 4-momentum $p_i$ (in a mixed $p,x$
representation) occurring in the structure of the vertex function. This
amounts to replacing each ${\hat q}_\mu$ occurring in $\Gamma({\hat q})$ =
$D({\hat q})\phi({\hat q})$, by ${\hat q}_\mu -e_q {\hat A}_\mu$, where
$e_q = {\hat m}_2 e_1 -{\hat m}_1 e_2$, and keeping only first order terms
in $A_\mu$ after due expansion. Now the first order correction to
${\hat q}^2$ is $-e_q{\hat q}.{\hat A}- e_q{\hat A}.{\hat q}$, which
simplifies on substitution from  eq.(2.1) to
\begin{equation}
-2e_q \tilde q.A \equiv -2e_q A_\mu [{\hat q}_\mu -{\hat q}.n {\tilde n}_\mu
+ P.{\tilde n}{\hat q}.n n_\mu /P.n]
\end{equation}
The net result is a first order correction to $\Gamma({\hat q})$ of amount
$e_q j({\hat q}).A$ where
\begin{equation}
j({\hat q})_\mu = -4M_>{\tilde q}_\mu\phi({\hat q}) (1-({\hat q}^2 -
\lambda/{4M^2})/{2\beta^2})
\end{equation}
The contribution to the P-meson form factor from this hadron-quark-photon
vertex (4-point) now gives the QED gauge correction to the triangle loops,
figs.(1a,1b) of [19], to the main term $F(k^2)$, eq.(4.1), of an amount
which, after a simple trace evaluation (and anticipating the vanishing
of all $\Delta$-terms remaining in the trace, as a result of contour
integration over $q_n$)  simplifies to ($\phi=\phi({\hat q})$, etc)
\begin{equation}
F_1(k^2) = 4(2\pi)^4 N_H^2 e_q {\hat m}_1 M_>^2
\int d^4q
(M^2-{\delta m}^2)\phi\phi'[{{D_n'{\tilde q}.{\bar P}} \over
{\Delta_1' \Delta_2 P'.n}} + {{D_n {\tilde q}'.{\bar P}} \over
{\Delta_1 \Delta_2 P.n}}]+ [1 \Rightarrow 2];
\end{equation}
In writing down this term, the proportionality of the current to
$2{\bar P}_\mu$ has been incorporated on both sides, on identical lines
to that of (4.1), using results from (4.2-4.7) as well as from Appendix A.
Note that $e_q$ is {\it antisymmetric} in `1' and `2', signifying a change
of sign when the second term $[1 \Rightarrow 2]$ is added to the first.
The term ${\tilde q}.{\bar P}/{\bar P}^2$ simplifies to
$2q.n (1-{\hat k}/2)/P.n$, after extracting the proportionality to
${\bar P}_\mu$. Next, after the pole integrations over $q_n, q_n'$ in
accordance with (4.7), it is useful to club together the results of photon
insertions on {\it both} blobs for either index (`1' or `2'); this step
generates two independent combinations for the `1' terms (and similarly
for `2' terms):
\begin{equation}
A_n = q.n(1-{\hat k}/2); \quad
B_n = q.n(1-{\hat k}/2)({\hat q}^2 -\lambda/{4M_>^2})/{2\beta^2}
\end{equation}
Collecting all these contributions the result of $q_n$- integration is
\begin{equation}
F_1(k^2)= 8(2\pi)^3 N_H^2 e_q {\hat m}_1 M_>^2 \int d^3{\hat q}
(M^2-{\delta m}^2)\phi\phi'[{{A_n +A_n'-B_n-B_n'} \over {\eta_k \times
({\bar P}.n)^2}}] + [1 \Rightarrow 2]
\end{equation}
The rest of the calculation is routine and follows closely the steps of
Appendix A for the (main) $F(k^2)$ term, including the translation
$z_2 \rightarrow z_2 + {\hat m}_2 {\hat k}^2/{2\theta_k}$, and is omitted
for brevity. The final result for $F_1(k^2)$ is
\begin{equation}
F_1(k^2) = -e_q{\hat m}_1{\hat m}_2 (3\eta_k+{\hat k}^2)[{{(M_>^2-{\delta
m}^2)
\eta_k{\hat k}^4}(M_> {\hat m}_2)^2
\over {8G(0) \theta_k^{7/2} \beta^2}}]  + [1 \Rightarrow 2]
\end{equation}
where we have dropped some terms which vanish on including the
$[1 \Rightarrow 2]$ terms, noting the $(1,2)$ antisymmetry of $e_q$.

\subsection{Large and Small $k^2$ Limits of Form Factor}
\par
        We close this section with the large and small $k^2$ limits of
the form factors $F(k^2)$ and $F_1(k^2)$. For large $k^2$, eq.(4.11) gives
${\hat k}$= 2, $\theta_k$ = 2, and $\eta_k$ = $4M^2/k^2$, so that
\begin{equation}
F(k^2)= 2M N_H^2 (2\pi)^3 {\hat m}_1 {{4M^2} \over k^2} (\pi
\beta^2/2)^{3/2}
G(\inf) exp[-{(M{\hat m}_2/\beta)^2/2}]   +[1 \Rightarrow 2]
\end{equation}
where, from eqs.(A.11-12),
\begin{equation}
G(\inf) = (1+{\delta m}^2/M^2)(\beta^2- \lambda/{4M^2}+M^2{\hat m}_2^2)
+(M^2-{\delta m}^2){\hat m}_2 -2{\hat m}_2^2 M^2
\end{equation}
Similarly from eq.(4.14), the large $k^2$ limit of $F_1(k^2)$ is
\begin{equation}
F_1(k^2) = 2{\sqrt 2}M^2k^{-2} e_q{\hat m}_1{\hat m}_2 (M_>^2-{\delta m}^2)
[{{M_>^2({\hat m}_1-{\hat m}_2)} \over {\beta^2 G(0)}}]
\end{equation}
where we have taken account of the $(1,2)$ antisymmetry of $e_q$ in
simplifying the effect of the $[1 \Rightarrow 2]$ term on the RHS. As a
check, both $F(k^2)$ and $F_1(k^2)$ are seen to satisfy the `scaling'
requirement of a $k^{-2}$ variation for large $k^2$. This result can be
traced to the input dynamics of the (non-perturbative) gluonic interaction,
eq.(3.2), on the structure of the vertex function, eq.(3.14). Perturbative
QCD of course gives a $k^{-2}$ behaviour [31]. The covariant NP approach of
KK [7] also gives a similar behaviour, but extracted in a somewhat
different way from the present `Lorentz completion' treatment. Note that
for the pion case the QED gauge correction term $F_1(k^2)$ gives zero
contribution in the large $k^2$ limit.
\par
        For small $k^2$, on the other hand, we have from eq.(4.11)
\begin{equation}
{\hat k}^2 = k^2/M^2; \quad [\theta_k, \eta_k] = 1 \pm  k^2/{4M^2}
\end{equation}
In this limit, the form factor, after substituting for $N_H$ from (4.8),
and summing over the `1' and `2' terms, works out as
\begin{equation}
F(k^2)= (1-3k^2/8M^2)[1- {\hat m}_1{\hat m}_2({{k^2} \over {4\beta^2}}-
{{k^2{\delta m}^2} \over {M^2 G(0)}}) -{{3 k^2 \beta^2 (1+{\delta m}^2/M^2)}
 \over {8M^2G(0)}}]
\end{equation}
where $G(0)$ is formally given by eq.(A.10), except for the replacement of
${\hat q}^2$ by $3\beta^2/2$. As a check, $F(k^2)$ is symmetrical in `1'
`2', as well as satisfies the consistency condition $F(0)$ = 1. Similarly
the small $k^2$ value of $F_1(k^2)$, after taking account of the $(1,2)$
antisymmetry of $e_q$, is of minimum order $k^4$, so that it contributes
neither to the normalization ($F_1(0)=0$), nor to the P-meson radius.
\par
        For completeness we record some numerical results for large and
small $k^2$ limits. For the pion case, in the large $k^2$ limit,
eqs.(4.12-13) yield after a little simplification the simple result
\begin{equation}
F(k^2) = C/k^2; \quad C \equiv  2{\sqrt 2}{M_>^2 \over G(0)} (\beta^2+m_q^2)
e^{-M_>^2/{8\beta^2}}
\end{equation}
where $m_q= 265 MeV$ stands for $m_1=m_2$; and $M_>$= $max{m_1+m_1, M}$.
Substituting for $\beta^2$=$0.0603 GeV^2$ [18] and $G(0)$=$0.166 GeV^2$,
yields the result $C$=$0.35 GeV^2$, vs the expt value of $0.50 \pm 0.10$
[28b]. For comparison, we also list the perturbative QCD value [31] of
$8\pi\alpha_s {f_\pi}^2$ = $0.296 GeV^2$, with $f_\pi$=$133MeV$, and the
argument $Q^2$ of $\alpha_s$ taken as $M_>^2$.
\par
        For low $k^2$, eqs.(4.14-15) yield values of the pion and kaon
radii, in accordance with the relation $<R^2>$ = $-{\nabla_k}^2F(k^2)$ in
the $k^2$=$0$ limit. Substitution of numerical values from (3.4-5) yields
\begin{equation}
R_K = 0.629 fm (vs: 0.53- expt [28a]); \quad
R_\pi = 0.661 fm (vs: 0.656- expt [28a])
\end{equation}

\section{Discussion, Summary and Conclusion}

In this paper we have tried to give an exposition of the Markov-Yukawa
Transversality Principle (MYTP) [8-10] as an alternative to the traditional
methods of 3D BSE reduction [1-3], vis-a-vis intrinsically 3D light-
front methods [4-7]. A basic difference between  the two approaches lies
in the former's capacity to provide an {\it interconnection} between the
3D and 4D BS wave functions [8], via the facility of a {\it reconstruction}
of the 4D form in terms of the 3D ingredients, while the latter gives at
most a one-way connection, viz., the 3D form as an integral over the
4D BS amplitude [7], but {\it {not vice-versa}}. Thus MYTP can do with the
normal 4D Feynman rules to govern the construction of various quark loop
integrals [8,16] where off-shell momenta are allowed their full play, while
the intrinsically 3D BSE/LF approaches [1-7] (with on-shell momenta) must
rely upon essentially 3D Feynman diagrams (with their special rules [7,23]).
\par
        To push this obvious advantage of MYTP in facilitating a two-tier
strategy ,viz., i) the 3D BSE form for hadron spectroscopy, and ii) the
reconstructed 4D BS vertex function for accessing various types of 4D
quark loop integrals, (while retaining formal covariance), we have
attempted in this paper to extend its scope from the initial  covariant
instantaneity ansatz (CIA) [8] (in company with the Pervushin group [9]),
to a much wider domain. Now the CIA [8] is explicitly Lorentz covariant, but
the hadron-momentum dependence of each $Hq{\bar q}$ vertex function causes
a `Lorentz mismatch' among these in an arbitrary quark loop integral, and
thus {\it induces} the appearance of time-like momentum components in the
(gaussian) wave functions involved, which results in `poorly defined'
4D integrals. (This pathology is reminiscient of a similar problem first
encountered by Feynman et al in the famous FKR paper [36]). One just
escapes this pathology in 2-quark loops [18], but it reappears in 3-quark
loops [21] and above. A possible remedy lies in making the covariant 3D
treatment less dependent (than in CIA) on the individual hadron momentum
frames. It is precisely to meet this objective that the present approach
of extending MYTP from the individual hadron frames to a more universal
null-plane frame was conceived, so as to eliminate such unwanted (time-like)
momentum components (responsible for the ill-defined loop integrals).
\par
        Looking back on this strategy, Sec 4 on the P-meson form factor
already shows that the idea has worked, except for the null-plane (NP)
orientation dependence. This is not a basically new result, for a simple-
minded, conventional NP approach [19,22] to BS dynamics had already produced
many results of this paper, both on spectroscopy [17] as well as on
transition amplitudes [19], but had been  criticized [20] on grounds of
`non-
covariance'. The present treatment with an explicit formulation of Markov-
Yukawa Transversality (MYTP) [10-8] on a {\it covariant} null plane (CNP),
hopefully, keeps both the advantages, since the 4D loop integrals, as Sec.4
shows on the form factor calculations, are not only perfectly well-defined,
but even a good part of the $n_\mu$ dependence has got eliminated in the
process of ${\hat q}$ integration, while the remaining NP orientation
dependence has been transferred to the external (hadron) 4-momenta. In this
regard the present approach is already in the company of a wider light-front
community [7, 23] which has also to contend with some $n_\mu$ dependence.
The solution we have offered to overcome this problem is a simple-minded
prescription of `Lorentz completion' wherein a `collinear frame' ansatz
$P_\perp.q_\perp = 0$ is lifted on the external hadron momenta $P,P'$ etc,
{\it after} doing the internal ${\hat q}$ integration, so as to yield an
explicitly Lorentz-invariant result. The prescription, though different
from KK [7], is nevertheless self-consistent, at least for 3-point hadron
vertices, (and amenable to extension to higher-point vertices provided the
latter can be expressed as a combination of simpler 3-point vertices).
\par
        In Sec 2 we have  also offered a detailed comparison of the
present method with the covariant Light Front (LF) approach of KK [7].
In particular, the angular condition [25] seems to be almost trivially
satisfied as seen from the rotational invariance of the 3D BSE structure
(Secs 2-3). And the manifest covariance of the present approach, with a
second NP variable ${\tilde n}_\mu$ which is a natural ``dual''  partner
of $n_\mu$ (without a counterpart in other light-front approaches
[7, 23]), has obviated the need for  3-vectors and/or Lorentz
transformations [7] to meet the same ends.
\par
        To summarise, the result of invoking MYTP on a covariant NP has been
two-fold: retaining the formal covariance of CIA [8,16], and avoiding the
appearance of time-like components inside the loop integrals. Further, its
general self-consistency is evidenced by its capacity to reproduce most of
the results of the old-fashioned NP approach [19,22] on the one hand, and
providing identical results to CIA [8,16] on the spectroscopy front [17] on
the other. The results on the P-meson form factor $F(k^2)$ are also on
expected lines, with `kinematical' gauge invariance satisfied explicitly,
and
the QED gauge correction $F_1(k^2)$ showing identical features. Both terms
exhibit the (expected) $k^{-2}$ behaviour at large $k^2$, while for the
(equal mass) pion case, the QED correction gives zero contribution. And
although the predicted value $0.35$ of the constant $C$ is somewhat less
than
the experimental value $0.50 \pm 0.1$ [28b], it is still not below the value
$8\pi \alpha_s f_\pi^2 = 0.296$ predicted by the QCD limit [31]. At the
opposite (small $k^2$) limit too, the e.m. radii of the pion and kaon are
in fair accord with experiment [28a]. Again the QED gauge correction does
not give any change up to order $k^2$. (We would like to add parenthetically
that the old-fashioned NP treatment [19] had yielded a slightly better curve
for the pion form factor, but this was due to the use of the ``half-off-
shell'' form of the NP wave function [19], which however did not come out
naturally from the present `covariant' treatment).
\par
        We would like to end with some remarks on the crucial role of the
Markov-Yukawa Transversality Principle [10] in  providing a natural access
to spectroscopy [11] as an integral part of any quark level study of hadron
physics, via the 3D sector of the 3D-4D BSE interconnection [8], while
retaining the full facility of 4D loop integrals via its 4D sector. Of
course
MYTP [10] by itself does not carry information on the `dynamics' of
spectroscopy which must be governed by other considerations (non-
perturbative QCD simulated by $DB\chi S$ [12-15]), but it at least
offers a broad enough framework to accommodate such dynamics, without having
to look elsewhere. We should also like to stress that the importance of
spectroscopy as an integral part of any `dynamical equation based' approach
merely reiterates  a philosophy initiated long ago by Feynman et al [36].

\section*{Appendix A: Derivation of $F(k^2)$ and $N_H$ for P-meson}
\setcounter{equation}{0}
\renewcommand{\theequation}{A.\arabic{equation}}

In this Appendix we outline the main steps to the derivation of the P-meson
form factor (4.9), as well as the Normalizer (4.8), given in Sec.4 of Text.
Collecting the various pieces after $p_{2n}$-pole integration, gives for
(4.1)
\begin{equation}
F(k^2) = 2(2\pi)^3 N_n(P)N_n(P'){\hat m}_1 \int d^2q_\perp dz_2 P.n
g(z_2)
e^{-q_\perp^2/\beta^2-f(z_2)/\beta^2} + [1 \Rightarrow 2];
\end{equation}
\begin{equation}
f(z_2)= M^2 \eta_k^{-2}[ \theta_kz_2^2-z_2{\hat k}^2{\hat m}_2
+\theta_k {\hat m}_2^2{\hat k}^2/4];
\end{equation}
\begin{equation}
 D_n+D_n'=4{\bar P}.n
[q_\perp^2+ M^2(z_2^2- z_2{\hat k}^2{\hat m}_2/2
+ {\hat m}_2^2 {\hat k}^2/4)/eta_k-\lambda/{4M^2}];
\end{equation}
\begin{equation}
 g(z_2)= {{D_n+D_n'} \over 4}{{M^2+{\delta m}^2} \over {M^2+k^2/4}}+h(z_2);
\end{equation}
\begin{equation}
h(z_2) = 2{\bar P}.n({\hat m}_2-z_2)[M^2-{\delta m}^2+
{\hat m}_2M^2({\delta m}^2-M^2-k^2/2)/(M^2+k^2/4)]
\end{equation}
The integration over $q_\perp$ and $z_2$ are both routine, the latter with
a translation $z_2 \rightarrow z_2+{1 \over 2}{\hat m}_2 {\hat
k}^2/\theta_k$,
to reduce the gaussian factor to the standard form. Note that, unlike the
conventional (Weinberg) form [4] of light-front dynamics, the present 4D
form
which permits off-shellness of the internal momenta, does not restrict
in principle the limits of $z_2$ integration. Thus after the translation,
the odd-$z_2$ terms can be dropped, and $f(z_2)$ reduces to
\begin{equation}
f(z_2)= M^2 z_2^2\theta_k/\eta_k^2+(M{\hat m}_2{\hat k})^2/{4\theta_k}
\end{equation}
while the $g$-function is a sum of two pieces $g_1+g_2$:
\begin{equation}
g_1 = \eta_k[q_\perp^2 +M^2z_2^2/\eta_k +
{1 \over 4} M^2 {\hat m}_2^2{\hat k}^2(1+3{\hat k}^2/4)/\theta_k^2
-\lambda/{4M^2}] (1+{\delta m}^2/M^2);
\end{equation}
\begin{equation}
g_2 = 2\eta_k (M^2-{\delta m}^2){\hat m}_2 /\theta_k +
2({\delta m}^2-M^2-k^2/2){\hat m}_2^2 {\eta_k}^2 /\theta_k
\end{equation}
\par
        Before writing the final result for $F(k^2)$, it is instructive at
this stage to infer the normalizer $N_H$ of the hadron, obtained by setting
$k_\mu=0$, and demanding that $F(0)=1$. This gives after some routine steps:
\begin{equation}
N_n(P)^{-2} = 2M (2\pi)^3 (P.n/ M)^2 \int d^3{\hat q}
e^{-{\hat q}^2/\beta^2} G(0);
\end{equation}
\begin{equation}
G(0) = [(1+{\delta m}^2/M^2)({\hat q}^2-\lambda/{4M^2})
+2{\hat m}_1{\hat m}_2 (M^2-{\delta m}^2)]
\end{equation}
where ${\hat q}=(q_\perp, Mz_2)$ is effectively a 3-vector, in conformity
with the requirements of the angular condition [7,23,25], which gives a
formal meaning to its third component $q_3$= $M q.n/P.n$ = $Mz_2$. The
normalization factor $N_n(P)$ is also seen to vary inversely as $P.n$,
while the multiplying integral is clearly independent of the NP-orientation
$n_\mu$. To exhibit this $P_n$ independence more explicitly, define a
`reduced normalizer' $N_H$ which equals $N_n(P) \times P.n/M$ and
gives for $N_H^{-2}$ the Lorentz-invariant result, eq.(4.8) of Text.
\par
        Now insert the result $N_n(P)$= $M N_H/P.n$ on the RHS of (A.1), and
note, via eq.(4.3), that
\begin{equation}
{M^2 /over {P.nP'.n}} = {M^2 /over {{{\bar P}.n}^2 \eta_k}}; \quad
\eta_k = 1-{\hat k}^2/4.
\end{equation}
One now checks that the factors ${\bar P}.n$ cancel out completely, and the
evaluation of the gaussian integrals leads after a modest algebra to
eq.(4.9)
of Text, where $G({\hat k})$, after collecting from eqs.(A.6-8), is given by
\begin{equation}
G({\hat k}) = (1+{\delta m}^2/M^2)h({\hat k})
+2(M^2-{\delta m}^2){\hat m}_2/\theta_k +2{\hat m}_2^2 \eta_k \theta_k^{-1}
({\delta m}^2-M^2-k^2/2);
\end{equation}
\begin{equation}
h({\hat k}) =  (1+{\eta_k}^2/{2\theta_k})\beta^2- \lambda/{4M^2}+
(M{\hat m}_2{\hat k}/2{\theta_k})^2(1+{3 \over 4}{\hat k}^2); \quad
{\delta m}= m_1-m_2.
\end{equation}

\end{document}